\documentclass[useAMS,usenatbib]{mn2e}
\usepackage{amsmath,amssymb}
\usepackage{delarray}
\usepackage{graphicx}
\usepackage{tabularx}

\newcommand{\Tv}{{\rm T_{vir}}}

\newcommand{\Msun}{\ensuremath{\mathrm{M}_{\odot}}}
\newcommand{\MDM}{{\rm M}}

\newcommand{\fcold}{\rm{f_{cold}}}

\newcommand{\zr}{{\rm z_{reion}}}
\newcommand{\RGC}{{\rm R_{GC}}}

\newcommand{\zcoll}{{\rm z_{coll}}}
\newcommand{\RI}{{\rm R_{I}}} 
\newcommand{\zI}{{\rm z_{I}}} 
\newcommand{\FO}{\rm F_0}
\newcommand{\tev}{\rm t_{ev}}
\newcommand{\zev}{\rm z_{ev}}
\newcommand{\zacc}{\rm z_{acc}}
\newcommand{\fesc}{\rm f_{esc}}
\newcommand{\Mt}{\rm M_{t}}
\newcommand{\zre}{\rm z_{re}}

\newcommand{\nicefrac}[2]{\leavevmode\kern.1em
            \raise.5ex\hbox{\the\scriptfont0 #1}\kern-.1em
      /\kern-.15em\lower.25ex\hbox{\the\scriptfont0 #2}}

\begin{document}


\title{A signature of the internal reionisation of the Milky Way?}



\author[Ocvirk \& Aubert]{Pierre\ Ocvirk$^{{1}}$ \&Dominique\ Aubert$^{{1}}$\\
$^1$Observatoire Astronomique de Strasbourg, 11 rue de l'Universit\'e, 67000 Strasbourg, France. \\
}


\date{Typeset \today; Received 15 April 2011 / Accepted 1 August 2011}

\pagerange{\pageref{firstpage}--\pageref{lastpage}} \pubyear{2011}

\maketitle
\label{firstpage}




\begin{abstract}
We present a new semi-analytical model of the population of satellite galaxies of the Milky Way, aimed at estimating the effect of the spatial structure of reionisation at galaxy scale on the properties of the satellites. In this model reionisation can be either: (A) externally-driven and uniform, or (B) internally-driven, by the most massive progenitor of the Milky-Way. In the latter scenario the propagation of the ionisation front and photon dilution introduce a delay in the photo-evaporation of the outer satellites' gas with respect to the inner satellites. As a consequence, outer satellites experience a longer period of star formation than those in the inner halo. We use simple models to account for star formation, the propagation of the ionisation front, photo-evaporation and observational biases. Both scenarios yield a model satellite population at z=0 that matches the observed luminosity function and mass-to-light ratios. However, the predicted population for scenario (B) is significantly more extended spatially than for scenario (A), by about 0.3 dex in distance, resulting in a much better match to the observations. 
The survival of this structural signature imprinted by the local UV field during reionisation on the radial distribution of satellites makes it a promising tool for studying the reionisation epoch at galaxy scale in the Milky Way and nearby galaxies resolved in stars with forthcoming large surveys.
However, more work is needed to determine how the effect reported here can be disentangled from that of cosmic variance between different realisations of Milky Way haloes.

\end{abstract}

\begin{keywords}
Galaxy: formation - galaxies: dwarf, luminosity function - Local Group - cosmology: reionization - diffuse radiation
\end{keywords}
%

\setcounter{page}{1}



\section{Introduction}


Reionisation took place as the first stars of the universe brought the dark ages to an end. 
It seems now well established that reionisation also had a profound impact on galaxy formation. The metagalactic UV radiation field may be responsible for evaporating the gas of low-mass galaxies \citep{gnedin2000,hoeft2006}, affecting their star formation and therefore also the buildup of the galactic halo \citep{bekkichiba2005}. This seems to provide an acceptable solution for the ``missing satellites problem'' \citep{klypin1999,moore1999}, by inhibiting star formation in low mass galaxies at early times \citep{bullock2000}.


Several recent semi-analytical models (SAMs) have shown that the observed luminosity function (hereinafter LF) of the satellites of the Milky Way (hereinafter MW) can be well reproduced under a number of simple assumptions for the baryonic physics \citep{koposov2009,munoz2009,busha2010,maccio2010,font2011}. They also reproduce the apparent common mass scale of $10^7 \Msun$ \citep{strigari2008}, although the latter is hotly debated \citep{aden2009,walker2010}. They suggest that the  ultra-faint dwarf galaxies (hereinafter UFDs) discovered by the SDSS \citep{martin2004,willman2005short,zucker2006,belokurov2007short, irwin2007short, walsh2007} are reionisation fossils, living in sub-haloes of about $10^{6-9} \Msun$.
The success of these models invites us to reconsider UFDs, not as a problem, but as probes of the reionisation epoch.


Most past studies implemented a uniform, instantaneous reionisation by simply suppressing star formation in the sub-haloes below $\zr$ \citep{koposov2009,busha2010,maccio2010}. Others tracked the evolution of the physical properties of the gas (temperature, neutral fraction) subject to the UV background from external and local sources, and SuperNova (SN) feedback, in order to determine self-consistently the satellite's reionisation redshift \citep{munoz2009,font2011}. \cite{lunnan2011} studied the effect of a complex, patchy reionisation on the satellites properties. In this letter, we investigate the effect of varying the structure of the UV field at reionisation on the properties of the satellites of the MW. We choose 2 deliberately simple configurations: a uniform background, and a unique, galacto-centric source. We show that the latter produces a significantly more extended satellite population.

We first present our model and its ingredients (Sec. \ref{s:model}), and then our results in terms of LF, masses, and radial distribution of satellites (Sec. \ref{s:results}). Our conclusions are presented in Sec. \ref{s:conclusion}.


\section{A semi-analytical model of the satellites of the Milky Way}
\label{s:model}
Our SAM is partly inspired by \cite{busha2010} and \cite{koposov2009} (hereinafter B10 and K09), but describes the reionisation epoch in more detail. We allow for a non-steady, non-uniform UV background and compute individual evaporation times for the gas hosted by the dark matter sub-haloes.

\subsection{Ingredients}
\label{s:ingredients}
The life of a model dark matter sub-halo in the range $10^{7-9}\Msun$ (typical of a MW satellite at z=0) can be summarized as follows: after its formation, it grows until it reaches a mass sufficient for forming stars. During reionisation, ionisation fronts (hereinafter I-front) of galactic or external origin propagate, and eventually overtake the sub-halo at a redshift $\zI$. The latter starts to lose its gas via photo-evaporation, which rapidly suppresses star formation, between 0.4-2 Gyr after the Big Bang. Then it undergoes a purely passive evolution for the next $\approx 6$ Gyr, until it is accreted by the MW, around $z\approx 1$. Depending on its impact parameters and concentration it will be tidally and ram-pressure stripped. Finally, it may get disrupted, losing all of its stars to the halo, or conserve a fraction of its pre-accretion mass and stellar content, and survive as a satellite until today. The purpose of the SAM described here is to model this chain of physical processes, with a number of deliberate simplifications, in order to produce a population of satellites and its observables. The major ingredients are listed below, in the order of the modelling steps.

\begin{enumerate}
\item{Sub-halo tracks: as in B10 we use the Via Lactea II (hereinafter VLII) data \citep{diemand2007}. The simulation represents a MW-size halo (about $2.10^{12} \Msun$) that has not had any recent major merger, making it a suitable host for a MW-like disk galaxy. The dark matter particle mass is $4.10^3 \Msun$, which allows about 20000 sub-haloes to be resolved, down to a few times $10^{5}\Msun$.}
\item{Threshold mass for star formation $\Mt$: a sub-halo is allowed to form stars when it reaches $\Tv=8.10^3$K, which translates to $\Mt \approx 10^{7-8} \Msun$. This reflects the typical halo mass where atomic hydrogen cooling becomes efficient, as in B10.}
\item{Following B10, once the threshold mass is reached, the halo forms stars at a rate 
${\rm SFR} = \epsilon ({\fcold \MDM_7})^{\alpha}$,
with $\epsilon = 10^{-5}$, $\fcold=\Omega_{\rm b}/\Omega_{\rm m}$, $\alpha =2$, and $\MDM_7=\MDM/10^7 \Msun$, where $\MDM$ is the dark matter mass of the sub-halo.
}
\item{Reionisation redshifts: we define the reionisation redshift $\zI$ as the time where the I-front {\em reaches} the sub-halo. The details for both external and internal scenarios are summarized in Sec. \ref{s:scenarios}. At variance with most of the literature, star formation does {\em not} end at $\zI$, as photo-evaporation is not considered instantaneous here.}
\item{Photo-evaporation: we compute evaporation times for the gas in the haloes using Eq. (5) of \cite{iliev2005}, 
as a function of the normalized flux $\FO$, the dark matter mass $\MDM$ of the halo, and its collapse redshift $\zcoll$: 
\begin{equation}
\tev = A  \MDM_{7}^{\rm B} \FO^{\rm C+D\,\log_{10}(\FO)} \left[ E + F \left( \frac{1+\zcoll}{10} \right) \right] {\rm Myr}  \, ,
\label{eq:tevap}
\end{equation}
where (A,B,C,D,E,F) are taken from \cite{iliev2005} for the case of a $5.10^{4}$ K black body.
Expectedly, $\tev$ increases with increasing mass and decreasing UV flux. 
For simplicity we assume a constant $\zcoll=15$ for all haloes. Photo-evaporation starts at $\zI$, and $\tev$ is converted to a $\Delta \zev$, so that evaporation finishes at a redshift $\zev =\zI + \Delta \zev$.
The SFR of the sub-halo is set to 0 for $z<\zev$. 
 The values taken by $\FO$ in our 2 reionisation scenarios are detailed in Sec. \ref{s:scenarios}.}
\item{The accretion redshift $\zacc$ is computed for each sub-halo using the VLII data. We consider that accretion on the MW halo results in instantaneous gas removal (if any left) from the satellite through ram-pressure stripping. We set ${\rm SFR}=0$ for $z<\zacc$. We can thus write the SFR$(\Msun/{\rm yr})$ prescription more synthetically as:


\begin{equation}
 \displaystyle
  {\rm SFR}= 
  \quad
  \left\{\begin{array}{l}
   \epsilon (\fcold \MDM_7 )^{\alpha} \mbox{ if}  
   \left\{\begin{array}{l}
   \Tv \geq  8.10^3 {\rm K} \mbox{ and}\\
   z  >  {\rm max}(\zev,\zacc) 
   \end{array} \right. \\
   0 \mbox{ otherwise} \, .
  \end{array}\right.
\label{eq:SFR}
\end{equation}
This formalism gives rise to typical periods of star formation of several 100 Myr between $z=15-6$, with SFR=$10^{-6}-10^{-2} \Msun/{\rm yr}$ for the observable satellites.
}
\item{Tidal stripping and disruption: A halo is considered disrupted if it loses more than 99\% of its pre-accretion mass. However, for a surviving halo we assume that the stellar content remains untouched. This is similar to the approach of K09 and \cite{maccio2010}.}
\item{Stellar population models: integrating eq. (\ref{eq:SFR}) over time yields the stellar mass and the individual star formation history (SFH) of the sub-haloes. We use the latter along with the \cite{BC03} models to assign luminosities to our satellite population, assuming a constant low metallicity Z=0.0004.}
\end{enumerate}

\subsection{Reionisation scenarios}
\label{s:scenarios}
We consider two scenarios, where reionisation of the IGM and the subsequent evaporation of the satellites' gas is driven either by an external or a galacto-centric source. Note that here the 2 scenarios are mutually exclusive.

\subsubsection*{{\bf Scenario (A): external reionisation}}
\label{s:external}
In this scenario, the UV background responsible for photo-evaporating the gas of the satellites is of extragalactic origin. We assume it is uniform, as in a majority of previous studies, and appears at $\zI=15$ for all sub-haloes, corresponding to the beginning of the reionisation of the IGM \citep{kogut2003}. We tune the value of the background flux to a constant $\FO = 0.03$, in the units of \cite{iliev2005}, so as to match the observed LF of the satellites. This translates into ${\rm J_{21}}=0.03$, which is compatible with the fluxes predicted by \cite{aubert2010} for the considered redshift range, although in the upper envelope.

\subsubsection*{{\bf Scenario (B): internal reionisation by a galacto-centric source}}

While reionisation appears patchy in large scale computations \citep{zahn2007}, the morphology of I-fronts at the scale of the MW progenitor (5 comoving Mpc) is rather unclear, and depends strongly on the Lyman continuum escape fraction $\fesc$ \citep{iliev2011}, which is currently very uncertain.
Here we assume that star formation within the most massive progenitor of the MW provides the photons responsible for evaporating the satellites' gas and the intervening IGM. This model results in a radially decreasing reionisation redshift profile in the IGM around the emitter, similar to those of Fig. 1 of \cite{lunnan2011}. However this model can not account for any stochasticity of the reionisation history of the IGM.

\begin{enumerate}
\item[(1) Initial I-front propagation:]{
we model the I-front $\RI(z)$ as a cosmological Str\"omgren sphere \citep{barkana2001} with a clumping factor C=1. The SFR of the galacto-centric source is an extrapolation from Fig. 4 of \cite{salvadori2010}: ${\rm SFR(z)=SFR_{25} 10^{(25-z)/10}}$, with ${\rm SFR_{25}=SFR(z=25)=0.02}\, \Msun$/yr.
We assume that a rate SFR=1\Msun/yr produces $1.6 \, 10^{53}$ ionising photons/s as in \cite{shapiro2004}.
We set the start of star formation at $z=25$, as suggested by theoretical studies \citep{bromm2009}.
Numerical simulations from \cite{razoumov2010} show that at z$>10$, $\fesc$ is close to 1. Therefore we use a constant $\fesc=1$ throughout the model.}
\item[(2) Photo-evaporation of the sub-haloes:]{
for a satellite at a distance $\RGC$ from the Galactic center, photo-evaporation starts at $\zI$, when the expanding I-front reaches the satellite, i.e. $\RI(\zI)=\RGC(\zI)$. We convert our SFR to $\FO$ using $\FO=16\,  \,SFR(z) \, ({10 \, {\rm kpc}}/{\RGC})^2$, following the conversions of \cite{shapiro2004}.}
\end{enumerate}

\subsection{Observational sample and biases}
We consider the same observed satellite population as in Tab. 2 of K09, consisting of the 11 classical and the 10 recently discovered SDSS satellites. The galacto-centric distances are taken from \cite{kroupa2010}. The observed luminosity function is corrected for the sky coverage of DR5 and the detection efficiency given by \cite{koposov2008}.
Detected model satellites are required to be within 260 kpc of the galactic center (the maximum distance where the tip of the red giant branch is detected by SDSS). Moreover they need to be within the completeness radius at a given magnitude as given by eq.(8) of K09 at z=0. Finally, their detection efficiency is computed using eq.(8) of \cite{koposov2008}.

\section{Results}
\label{s:results}
The LF is degenerated with respect to the parameters $(\epsilon,\FO)$: a decrease in star formation efficiency $\epsilon$ can be compensated by a longer star formation period, allowed by a smaller UV background flux. This is similar to the $\epsilon-\zre$ degeneracy reported in B10. In the present paper, $\FO$ was set so as to achieve evaporation of the satellites at high redshift { on average}, i.e. $<\zev>\approx 6$. 

\subsection{Luminosity function}
\begin{figure}
  {\includegraphics[width=0.95\linewidth,clip]{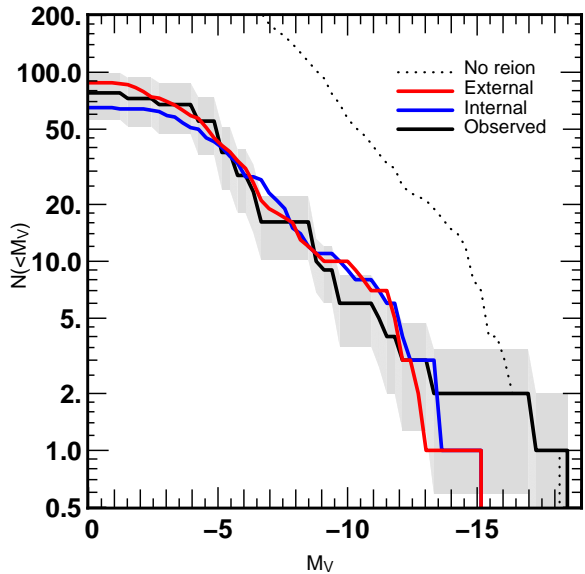}}
  \caption{Observed and model cumulative luminosity functions. The shaded areas show the observational error bars corresponding to Poisson noise, corrected for biases for the SDSS satellites. Both models with reionisation provide a good fit to the observations.}
\label{f:LF}
\end{figure}

Model and observed cumulative LFs are compared in Fig. \ref{f:LF}. 
Both models with reionisation agree equally well the observations. At he faint end, they differ slightly, but are both equally likely given the available data. At the bright end we underestimate the luminosity of the satellites, because of missing physics for the high-mass objects. For instance, MW's most massive satellites, such as the large and small Magellanic clouds, are still forming stars at z=0, which is forbidden in our model due to the shutdown of star formation subsequent to accretion on the MW halo and ram-pressure stripping. Our internal and external reionisation scenarios are indistinguishable when considering the existing data on the LF alone. To illustrate the overall effect of reionisation, we also show the results of a model identical to model (A) but without reionisation. It overestimates the abundance of satellites by a factor 10 for all but the 2 brightest objects and is therefore not favored in our framework. We refer the reader to \cite{li2010} for an in-depth exploration of models without reionisation.


\subsection{Masses}
\begin{figure}
  {\includegraphics[width=1.11\linewidth,clip]{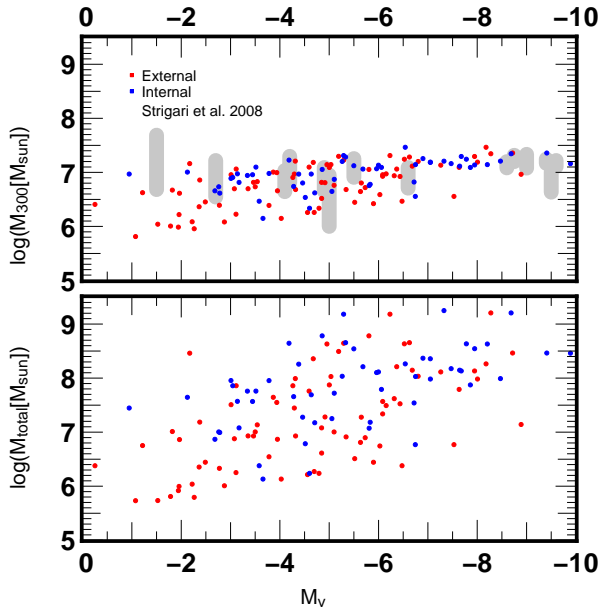}}
  \caption{Masses of the faint satellites. {\em Top:} central masses (${\rm M_{300}}$), models and as derived by Strigari et al. (2008). {\em Bottom:} total masses of simulated satellites. 
}
\label{f:m300}
\end{figure}
The central masses ($\rm M_{300}$, within the 300 central pc) of the model satellites are compared with observations in the top panel of Fig. \ref{f:m300}. They are taken directly from the VLII data at z=0. 
Both reionisation scenarios reproduce the range of values of ${\rm M_{300}}$ and $\rm{M_V}$ derived by \cite{strigari2008}. The distribution of ${\rm M_{300}}$ is almost flat, but there is a slight ${\rm M_{300}} - \rm{M_V}$ correlation which is not clearly seen in the data. This is typical of this class of models and is seen in all published works so far (\cite{munoz2009,maccio2010,font2011};K09;B10), as noted by \cite{kroupa2010}. 
Eventhough the physical meaning of this relation is currently unclear \citep{walker2010}, we include it as a comparison with previously published models.
The bottom panel of Fig. \ref{f:m300} shows the total dark matter masses of the satellites: despite the apparent common mass scale of ${\rm M_{300}} \approx 10^7 \Msun$, the satellites do indeed span a wide range in {\em total} mass, with a large scatter.
As for the LF, the masses of the satellites do not allow discrimination between external and internal reionisation.

\subsection{Radial distribution of satellites}

\begin{figure}
{\includegraphics[width=0.99\linewidth,clip]{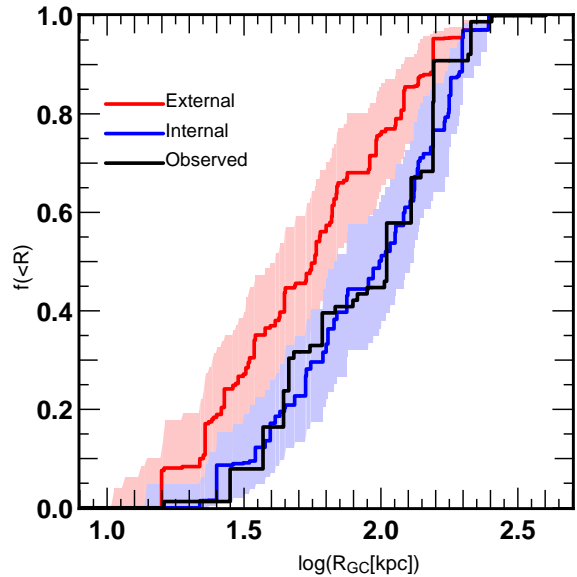}}
  \caption{Cumulative radial distribution profiles, observed (black), and predicted by the internal and external reionisation models at z=0. The shaded areas show the dispersion of profiles obtained by 1000 realisations of a mock DR5 survey. For better readability, the blue area is plotted on top of the red one, thereby partially masking it at the top and bottom of the figure.
}
\label{f:crd}
\end{figure}

In order to quantify the impact of the limited sky coverage of the SDSS (about 1/5 in area), we produced  mock surveys using our model satellite population: we include the 11 brightest to simulate the ``classical'' satellites and randomly pick 10 of the fainter ones within the detectable subsample (i.e. the ones that make up the LFs of Fig. \ref{f:LF}). We show the mean (solid lines) and dispersion (shaded areas) of the cumulative radial distributions at z=0 obtained for 1000 of these mock surveys in Fig. \ref{f:crd}. There is a strong difference in the predicted profiles of our two reionisation scenarios:  the distribution for the internal reionisation model is shifted outwards by about 0.3 dex with respect to the external reionisation model. 
This shift is 2 times larger than the dispersion of the predicted profiles, and is therefore significant. 
This is the signature of internal reionisation: in the galacto-centric reionisation scenario the satellites of the inner halo see a more intense UV flux than their outer halo cousins, and therefore evaporate faster. Thus the outer halo satellites experience a longer star formation activity period and  end up brighter. 
In this model, all the detected satellites beyond R$>$150 kpc see their photo-evaporation achieved later than z=6, the epoch of end of reionisation of the IGM derived by \cite{Fan2006}. 

\section{Conclusions and discussion}
\label{s:conclusion}

We developed a SAM of the satellite population of the MW based on the VLII simulation, simple recipes for star formation and radiative effects, and taking into account the observational biases. Our treatment of the propagation of the I-front within the early MW progenitor and photo-evaporation allows us to explore the effect of the structure of the UV field during reionisation on the properties of the satellite population.
In the galacto-centric reionisation model, the satellites of the inner halo are exposed earlier to a nearer UV source, and therefore evaporate faster than the outer halo satellites. As a consequence star formation in the inner halo is suppressed sooner, whereas in the outer halo, satellites continue forming stars until up to several 100 Myr later. The resulting radial distribution of satellites is shifted outwards by 0.3 dex in radius with respect to the external reionisation model, and provides a very good match to the observations. 

The sensitivity of the radial distribution of satellites to the structure of the UV field during reionisation makes it an interesting probe for the study of reionisation at galaxy scale in the MW with Pan-STARRS and upcoming LSST.
Provided that the VLII halo is representative of the MW, our result suggests that star formation at the center of our Galaxy is responsible for the photo-evaporation of its own satellites and the intervening gas.
However this interpretation is complicated by cosmic variance in the assembly history of the MW, which can also shift the distribution, as shows Fig. 14 of K09. In order to assess the significance of our result for the reionisation history of the MW, we will need to couple our model to  a larger set of high resolution simulations of the formation of the MW similar to the VLII, such as the AQUARIUS dataset \citep{aquarius}, in the spirit of \cite{font2011,lunnan2011}.





Moreover, a number of simplifications were made in this work, which need to be improved on. Firstly, the \cite{iliev2005} photo-evaporation simulations were designed to be mostly reliable around $10^7 \Msun$. Thus, we need to check via further numerical simulations that the evaporation times we extrapolated up to the $>10^{8} \Msun$ regime remain realistic. We will also investigate the effect of endogenous reionisation (i.e. induced by stars formed within the dwarf galaxy itself), which has been neglected in this work. 
Secondly, we need to generalize the present framework to account for multiple sources in the internal scenario: some fraction of the less massive progenitor haloes may also contribute to the UV field. Even though they are clustered around the most massive progenitor, they could lead to a more patchy UV field around the proto-MW, as seen at larger scale in \cite{zahn2007}. The transient nature of the radiation field resulting from the short lifetime of massive stars is another untested aspect of the model. Accurate evaluation and modelling of these effects is crucial for the interpretation of Pan-STARRS and LSST surveys of the satellite galaxies of the MW. Since the latter are major progenitors of MW stellar halo and streams, as illustrated in \cite{gomez2010}, accurate interpretation of Gaia data 
will render this detailed understanding of the reionisation of dwarf galaxies mandatory.

\section*{Acknowledgements}
PO is grateful to B. Famaey, C. Boily, R. Ibata, M. Mouhcine, J. Chardin for stimulating discussions in the course of this study. 
This work was carried out in the context of the LIDAU (LIght in the Dark Ages of the Universe) project, which is funded by the french ANR (Agence Nationale pour la Recherche, project code ANR-09-BLAN-0030). Additional thanks go to all LIDAU members and workshops participants for providing useful criticisms of the model and a motivating work atmosphere, to M. Busha for communicating details of his models, and J. Diemand for making the Via Lactea II data public.
PO also thanks D.~Munro for freely distributing his Yorick programming language (available at \texttt{http://yorick.sourceforge.net/}), which was used throughout these investigations.

\bibliographystyle{mn2e}
\bibliography{mybib}

\begin{thebibliography}{}

\bibitem[\protect\citeauthoryear{{Ad{\'e}n}, {Wilkinson}, {Read}, {Feltzing},
  {Koch}, {Gilmore}, {Grebel} \& {Lundstr{\"o}m}}{{Ad{\'e}n}
  et~al.}{2009}]{aden2009}
{Ad{\'e}n} D.,  {Wilkinson} M.~I.,  {Read} J.~I.,  {Feltzing} S.,  {Koch} A.,
  {Gilmore} G.~F.,  {Grebel} E.~K.,    {Lundstr{\"o}m} I.,  2009, \apjl, 706,
  L150

\bibitem[\protect\citeauthoryear{{Aubert} \& {Teyssier}}{{Aubert} \&
  {Teyssier}}{2010}]{aubert2010}
{Aubert} D.,  {Teyssier} R.,  2010, \apj, 724, 244

\bibitem[\protect\citeauthoryear{{Barkana} \& {Loeb}}{{Barkana} \&
  {Loeb}}{2001}]{barkana2001}
{Barkana} R.,  {Loeb} A.,  2001, \physrep, 349, 125

\bibitem[\protect\citeauthoryear{{Bekki} \& {Chiba}}{{Bekki} \&
  {Chiba}}{2005}]{bekkichiba2005}
{Bekki} K.,  {Chiba} M.,  2005, \apjl, 625, L107

\bibitem[\protect\citeauthoryear{{Belokurov} \& {et al.}}{{Belokurov} \& {et
  al.}}{2007}]{belokurov2007short}
{Belokurov} V.,  {et al.} 2007, \apj, 654, 897

\bibitem[\protect\citeauthoryear{{Bromm}, {Yoshida}, {Hernquist} \&
  {McKee}}{{Bromm} et~al.}{2009}]{bromm2009}
{Bromm} V.,  {Yoshida} N.,  {Hernquist} L.,    {McKee} C.~F.,  2009, \nat, 459,
  49

\bibitem[\protect\citeauthoryear{{Bruzual} \& {Charlot}}{{Bruzual} \&
  {Charlot}}{2003}]{BC03}
{Bruzual} G.,  {Charlot} S.,  2003, \mnras, 344, 1000

\bibitem[\protect\citeauthoryear{{Bullock}, {Kravtsov} \& {Weinberg}}{{Bullock}
  et~al.}{2000}]{bullock2000}
{Bullock} J.~S.,  {Kravtsov} A.~V.,    {Weinberg} D.~H.,  2000, \apj, 539, 517

\bibitem[\protect\citeauthoryear{{Busha}, {Alvarez}, {Wechsler}, {Abel} \&
  {Strigari}}{{Busha} et~al.}{2010}]{busha2010}
{Busha} M.~T.,  {Alvarez} M.~A.,  {Wechsler} R.~H.,  {Abel} T.,    {Strigari}
  L.~E.,  2010, \apj, 710, 408

\bibitem[\protect\citeauthoryear{{Diemand}, {Kuhlen} \& {Madau}}{{Diemand}
  et~al.}{2007}]{diemand2007}
{Diemand} J.,  {Kuhlen} M.,    {Madau} P.,  2007, \apj, 667, 859

\bibitem[\protect\citeauthoryear{{Fan}, {Strauss}, {Becker}, {White}, {Gunn},
  {Knapp}, {Richards}, {Schneider}, {Brinkmann} \& {Fukugita}}{{Fan}
  et~al.}{2006}]{Fan2006}
{Fan} X.,  {Strauss} M.~A.,  {Becker} R.~H.,  {White} R.~L.,  {Gunn} J.~E.,
  {Knapp} G.~R.,  {Richards} G.~T.,  {Schneider} D.~P.,  {Brinkmann} J.,
  {Fukugita} M.,  2006, \aj, 132, 117

\bibitem[\protect\citeauthoryear{{Font}, {Benson}, {Bower}, {Frenk}, {Cooper},
  {De Lucia}, {Helly}, {Helmi}, {Li}, {McCarthy}, {Navarro}, {Springel},
  {Starkenburg} \& {Wang}}{{Font} et~al.}{2011}]{font2011}
{Font} A.~S.,  {Benson} A.~J.,  {Bower} R.~G.,  {Frenk} C.~F.,  {Cooper} A.~P.,
   {De Lucia} G.,  {Helly} J.~C.,  {Helmi} A.,  {Li} Y.,  {McCarthy} I.~G.,
  {Navarro} J.~F.,  {Springel} V.,  {Starkenburg} E.,    {Wang} J.,  2011,
  ArXiv e-prints

\bibitem[\protect\citeauthoryear{{Gnedin}}{{Gnedin}}{2000}]{gnedin2000}
{Gnedin} N.~Y.,  2000, \apj, 542, 535

\bibitem[\protect\citeauthoryear{{G{\'o}mez}, {Helmi}, {Brown} \&
  {Li}}{{G{\'o}mez} et~al.}{2010}]{gomez2010}
{G{\'o}mez} F.~A.,  {Helmi} A.,  {Brown} A.~G.~A.,    {Li} Y.,  2010, \mnras,
  408, 935

\bibitem[\protect\citeauthoryear{{Hoeft}, {Yepes}, {Gottl{\"o}ber} \&
  {Springel}}{{Hoeft} et~al.}{2006}]{hoeft2006}
{Hoeft} M.,  {Yepes} G.,  {Gottl{\"o}ber} S.,    {Springel} V.,  2006, \mnras,
  371, 401

\bibitem[\protect\citeauthoryear{{Iliev}, {Moore}, {Gottl{\"o}ber}, {Yepes},
  {Hoffman} \& {Mellema}}{{Iliev} et~al.}{2011}]{iliev2011}
{Iliev} I.~T.,  {Moore} B.,  {Gottl{\"o}ber} S.,  {Yepes} G.,  {Hoffman} Y.,
  {Mellema} G.,  2011, \mnras, pp 296--+

\bibitem[\protect\citeauthoryear{{Iliev}, {Shapiro} \& {Raga}}{{Iliev}
  et~al.}{2005}]{iliev2005}
{Iliev} I.~T.,  {Shapiro} P.~R.,    {Raga} A.~C.,  2005, \mnras, 361, 405

\bibitem[\protect\citeauthoryear{{Irwin} \& {et al.}}{{Irwin} \& {et
  al.}}{2007}]{irwin2007short}
{Irwin} M.~J.,  {et al.} 2007, \apjl, 656, L13

\bibitem[\protect\citeauthoryear{{Klypin}, {Kravtsov}, {Valenzuela} \&
  {Prada}}{{Klypin} et~al.}{1999}]{klypin1999}
{Klypin} A.,  {Kravtsov} A.~V.,  {Valenzuela} O.,    {Prada} F.,  1999, \apj,
  522, 82

\bibitem[\protect\citeauthoryear{{Kogut}, {Spergel}, {Barnes}, {Bennett},
  {Halpern}, {Hinshaw}, {Jarosik}, {Limon}, {Meyer}, {Page}, {Tucker},
  {Wollack} \& {Wright}}{{Kogut} et~al.}{2003}]{kogut2003}
{Kogut} A.,  {Spergel} D.~N.,  {Barnes} C.,  {Bennett} C.~L.,  {Halpern} M.,
  {Hinshaw} G.,  {Jarosik} N.,  {Limon} M.,  {Meyer} S.~S.,  {Page} L.,
  {Tucker} G.~S.,  {Wollack} E.,    {Wright} E.~L.,  2003, \apjs, 148, 161

\bibitem[\protect\citeauthoryear{{Koposov}, {Belokurov}, {Evans}, {Hewett},
  {Irwin}, {Gilmore}, {Zucker}, {Rix}, {Fellhauer}, {Bell} \&
  {Glushkova}}{{Koposov} et~al.}{2008}]{koposov2008}
{Koposov} S.,  {Belokurov} V.,  {Evans} N.~W.,  {Hewett} P.~C.,  {Irwin} M.~J.,
   {Gilmore} G.,  {Zucker} D.~B.,  {Rix} H.,  {Fellhauer} M.,  {Bell} E.~F.,
  {Glushkova} E.~V.,  2008, \apj, 686, 279

\bibitem[\protect\citeauthoryear{{Koposov}, {Yoo}, {Rix}, {Weinberg},
  {Macci{\`o}} \& {Escud{\'e}}}{{Koposov} et~al.}{2009}]{koposov2009}
{Koposov} S.~E.,  {Yoo} J.,  {Rix} H.,  {Weinberg} D.~H.,  {Macci{\`o}} A.~V.,
    {Escud{\'e}} J.~M.,  2009, \apj, 696, 2179

\bibitem[\protect\citeauthoryear{{Kroupa}, {Famaey}, {de Boer},
  {Dabringhausen}, {Pawlowski}, {Boily}, {Jerjen}, {Forbes}, {Hensler} \&
  {Metz}}{{Kroupa} et~al.}{2010}]{kroupa2010}
{Kroupa} P.,  {Famaey} B.,  {de Boer} K.~S.,  {Dabringhausen} J.,  {Pawlowski}
  M.~S.,  {Boily} C.~M.,  {Jerjen} H.,  {Forbes} D.,  {Hensler} G.,    {Metz}
  M.,  2010, \aap, 523, A32+

\bibitem[\protect\citeauthoryear{{Li}, {De Lucia} \& {Helmi}}{{Li}
  et~al.}{2010}]{li2010}
{Li} Y.,  {De Lucia} G.,    {Helmi} A.,  2010, \mnras, 401, 2036

\bibitem[\protect\citeauthoryear{{Lunnan}, {Vogelsberger}, {Frebel},
  {Hernquist}, {Lidz} \& {Boylan-Kolchin}}{{Lunnan} et~al.}{2011}]{lunnan2011}
{Lunnan} R.,  {Vogelsberger} M.,  {Frebel} A.,  {Hernquist} L.,  {Lidz} A.,
  {Boylan-Kolchin} M.,  2011, ArXiv e-prints

\bibitem[\protect\citeauthoryear{{Macci{\`o}}, {Kang}, {Fontanot},
  {Somerville}, {Koposov} \& {Monaco}}{{Macci{\`o}} et~al.}{2010}]{maccio2010}
{Macci{\`o}} A.~V.,  {Kang} X.,  {Fontanot} F.,  {Somerville} R.~S.,  {Koposov}
  S.,    {Monaco} P.,  2010, \mnras, 402, 1995

\bibitem[\protect\citeauthoryear{{Martin}, {Ibata}, {Bellazzini}, {Irwin},
  {Lewis} \& {Dehnen}}{{Martin} et~al.}{2004}]{martin2004}
{Martin} N.~F.,  {Ibata} R.~A.,  {Bellazzini} M.,  {Irwin} M.~J.,  {Lewis}
  G.~F.,    {Dehnen} W.,  2004, \mnras, 348, 12

\bibitem[\protect\citeauthoryear{{Moore}, {Ghigna}, {Governato}, {Lake},
  {Quinn}, {Stadel} \& {Tozzi}}{{Moore} et~al.}{1999}]{moore1999}
{Moore} B.,  {Ghigna} S.,  {Governato} F.,  {Lake} G.,  {Quinn} T.,  {Stadel}
  J.,    {Tozzi} P.,  1999, \apjl, 524, L19

\bibitem[\protect\citeauthoryear{{Mu{\~n}oz}, {Madau}, {Loeb} \&
  {Diemand}}{{Mu{\~n}oz} et~al.}{2009}]{munoz2009}
{Mu{\~n}oz} J.~A.,  {Madau} P.,  {Loeb} A.,    {Diemand} J.,  2009, \mnras,
  400, 1593

\bibitem[\protect\citeauthoryear{{Razoumov} \& {Sommer-Larsen}}{{Razoumov} \&
  {Sommer-Larsen}}{2010}]{razoumov2010}
{Razoumov} A.~O.,  {Sommer-Larsen} J.,  2010, \apj, 710, 1239

\bibitem[\protect\citeauthoryear{{Salvadori}, {Dayal} \& {Ferrara}}{{Salvadori}
  et~al.}{2010}]{salvadori2010}
{Salvadori} S.,  {Dayal} P.,    {Ferrara} A.,  2010, \mnras, 407, L1

\bibitem[\protect\citeauthoryear{{Shapiro}, {Iliev} \& {Raga}}{{Shapiro}
  et~al.}{2004}]{shapiro2004}
{Shapiro} P.~R.,  {Iliev} I.~T.,    {Raga} A.~C.,  2004, \mnras, 348, 753

\bibitem[\protect\citeauthoryear{{Springel}, {Wang}, {Vogelsberger}, {Ludlow},
  {Jenkins}, {Helmi}, {Navarro}, {Frenk} \& {White}}{{Springel}
  et~al.}{2008}]{aquarius}
{Springel} V.,  {Wang} J.,  {Vogelsberger} M.,  {Ludlow} A.,  {Jenkins} A.,
  {Helmi} A.,  {Navarro} J.~F.,  {Frenk} C.~S.,    {White} S.~D.~M.,  2008,
  \mnras, 391, 1685

\bibitem[\protect\citeauthoryear{{Strigari}, {Bullock}, {Kaplinghat}, {Simon},
  {Geha}, {Willman} \& {Walker}}{{Strigari} et~al.}{2008}]{strigari2008}
{Strigari} L.~E.,  {Bullock} J.~S.,  {Kaplinghat} M.,  {Simon} J.~D.,  {Geha}
  M.,  {Willman} B.,    {Walker} M.~G.,  2008, \nat, 454, 1096

\bibitem[\protect\citeauthoryear{{Walker}, {McGaugh}, {Mateo}, {Olszewski} \&
  {Kuzio de Naray}}{{Walker} et~al.}{2010}]{walker2010}
{Walker} M.~G.,  {McGaugh} S.~S.,  {Mateo} M.,  {Olszewski} E.~W.,    {Kuzio de
  Naray} R.,  2010, \apjl, 717, L87

\bibitem[\protect\citeauthoryear{{Walsh}, {Jerjen} \& {Willman}}{{Walsh}
  et~al.}{2007}]{walsh2007}
{Walsh} S.~M.,  {Jerjen} H.,    {Willman} B.,  2007, \apjl, 662, L83

\bibitem[\protect\citeauthoryear{{Willman} \& {et al.}}{{Willman} \& {et
  al.}}{2005}]{willman2005short}
{Willman} B.,  {et al.} 2005, \apjl, 626, L85

\bibitem[\protect\citeauthoryear{{Zahn}, {Lidz}, {McQuinn}, {Dutta},
  {Hernquist}, {Zaldarriaga} \& {Furlanetto}}{{Zahn} et~al.}{2007}]{zahn2007}
{Zahn} O.,  {Lidz} A.,  {McQuinn} M.,  {Dutta} S.,  {Hernquist} L.,
  {Zaldarriaga} M.,    {Furlanetto} S.~R.,  2007, \apj, 654, 12

\bibitem[\protect\citeauthoryear{{Zucker}, {Belokurov}, {Evans}, {Wilkinson},
  {Irwin}, {Sivarani}, {Hodgkin}, {Bramich}, {Irwin}, {Gilmore}, {Willman} \&
  {Vidrih}}{{Zucker} et~al.}{2006}]{zucker2006}
{Zucker} D.~B.,  {Belokurov} V.,  {Evans} N.~W.,  {Wilkinson} M.~I.,  {Irwin}
  M.~J.,  {Sivarani} T.,  {Hodgkin} S.,  {Bramich} D.~M.,  {Irwin} J.~M.,
  {Gilmore} G.,  {Willman} B.,    {Vidrih} S.,  2006, \apjl, 643, L103

\end{thebibliography}
%

\label{lastpage}
\end{document}